\documentclass[prl,twocolumn,showpacs]{revtex4}
\usepackage{graphicx}

\begin{document}

\newcommand{\ket}[1]{\mbox{$|\!#1\;\!\rangle$}}
\newcommand{\aver}[1]{\mbox{$<\!#1\!\!>$}}
\def\ua{\uparrow}
\def\da{\downarrow}

\title{Single-shot read-out of electron spin states in a quantum dot\\ using spin-dependent tunnel rates}

\author{R. Hanson, L. H. Willems van Beveren, I. T. Vink, J. M. Elzerman, W. J. M. Naber,\\ F. H. L. Koppens, L. P. Kouwenhoven, and L. M. K. Vandersypen}
\affiliation{Kavli Institute of Nanoscience, Delft University of Technology,\\ PO Box 5046, 2600 GA Delft, The Netherlands}

\date{\today}

\begin{abstract}
We present a method for reading out the spin state of electrons in a quantum dot that is robust against charge noise and can be used even when the electron temperature exceeds the energy splitting between the states. The spin states are first correlated to different charge states using a spin dependence of the tunnel rates. A subsequent fast measurement of the charge on the dot then reveals the original spin state. We experimentally demonstrate the method by performing read-out of the two-electron spin states, achieving a single-shot visibility of more than 80\%. We find very long triplet-to-singlet relaxation times (up to several milliseconds), with a strong dependence on in-plane magnetic field.
\end{abstract}


\maketitle

The spin of electrons in semiconductors is the subject of extensive research, partly motivated by the prospects of using the spin as a classical bit \cite{Wolf} or a quantum bit \cite{LossDiVincenzo}. Electron spins can be conveniently studied when confined to a semiconductor quantum dot,
since here the number of electrons can be precisely controlled (down to zero) \cite{Ciorga,JeroFewEl}, the tunnel coupling to the reservoir is tunable over a wide range \cite{JeroFewEl} and single-electron tunneling can be monitored in real-time using a nearby quantum point contact (QPC) \cite{EnsslinAPL,LievenAPL} or a single-electron transistor \cite{LuNature, FujisawaRealTimeAPL} as an electrometer.
For applications in quantum computing as well as for fundamental research such as a measurement of Bell's inequalities, it is essential that the spin state of the electrons can be read out.

\begin{figure}[!t]
\includegraphics[width=3.4in]{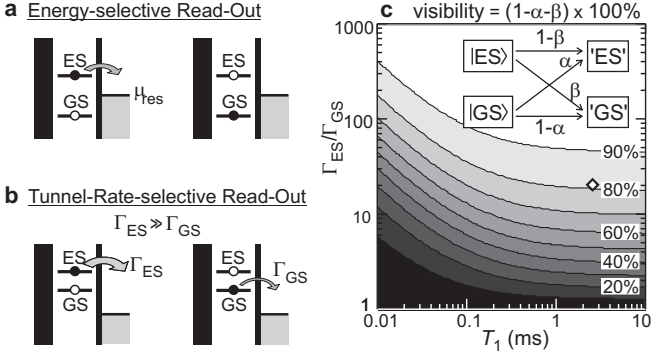}
\caption{(a)-(b) Energy diagrams explaining two schemes for spin-to-charge conversion.
(a) Energy-selective read-out. Tunneling is energetically allowed from \ket{\:ES} (left diagram), but not from \ket{\:GS} (right diagram).
(b) Tunnel rate-selective read-out (TR-RO). One electron is allowed to tunnel off the dot, regardless of the spin state, but the tunnel rate depends strongly on the spin state: $\Gamma_{ES}\!\gg\!\Gamma_{GS}$.
If a charge measurement after a time $\tau$, where $\Gamma_{GS}^{-1}\!\gg\!\tau\!\gg\!\Gamma_{ES}^{-1}$, indicates that one electron has (not) tunneled, the state is declared $'ES'$ ($'GS'$). 
(c) Visibility of the TR-RO as a function of spin relaxation time $T_1$ and the ratio $\Gamma_{ES}/\Gamma_{GS}$, for $\Gamma_{GS}$~=~2.5~kHz. The diamond corresponds to the read-out parameters of Fig. \ref{TR-ROFig2}e. Inset: definition of the error rates $\alpha$ and $\beta$. If the initial state is \ket{\:GS}, there is a probability $\alpha$ that the measurement gives the wrong outcome, i.e. $'ES'$ ($\beta$ is defined similarly).
}
\label{TR-ROFig1}
\end{figure}

The magnetic moment associated with the electron spin is tiny and therefore hard to measure directly. However, by correlating the spin states to different charge states and subsequently measuring the charge on the dot, the spin state can be determined \cite{LossDiVincenzo}. Such a spin-to-charge conversion can be achieved by positioning the spin levels around the electrochemical potential of the reservoir $\mu_{res}$ as depicted in Fig.~\ref{TR-ROFig1}a, such that one electron can tunnel off the dot from the spin excited state, \ket{\:ES}, whereas tunneling from the ground state, \ket{\:GS}, is energetically forbidden. By combining this scheme with a fast (40 kHz bandwidth) measurement of the charge dynamics, we have recently performed read-out of the spin orientation of a single electron, with a single-shot visibility up to 65\% \cite{NatureReadout}. (A conceptionally similar scheme has also allowed single-shot read-out of a superconducting charge qubit~\cite{Astafiev04}).
However, this energy-selective read-out (E-RO) has three drawbacks: (i) E-RO requires an energy splitting of the spin states larger than the thermal energy of the electrons in the reservoir. Thus, for a single spin the read-out is only effective at very low electron temperature and high magnetic fields (8 T and higher in Ref.~\cite{NatureReadout}). Also, interesting effects occurring close to degeneracy, e.g. near the singlet-triplet crossing for two electrons \cite{VitalyST}, can not be probed. (ii) Since the E-RO relies on precise positioning of the spin levels with respect to the reservoir, it is very sensitive to fluctuations in the electrostatic potential. Background charge fluctuations \cite{FujisawaChargeNoise}, active even in today's most stable devices, can easily push the levels out of the read-out configuration. (iii) High-frequency noise can spoil the E-RO by inducing photon-assisted tunneling from the spin ground state to the reservoir. Since the QPC is a source of shot noise, this limits the current through the QPC and thereby the bandwidth of the charge detection \cite{LievenAPL}. A different read-out method is desired that does not suffer from these constraints.

In this work, we present a spin read-out scheme where spin-to-charge conversion is achieved by exploiting the difference in \textit{tunnel rates} of the different spin states to the reservoir~\cite{EngelPRL04}. We outline the concept of this tunnel-rate selective read-out (TR$-$RO) in Fig.~\ref{TR-ROFig1}b. Assume that the tunnel rate from \ket{\:ES} to the reservoir, $\Gamma_{ES}$, is much higher than the tunnel rate from \ket{\:GS}, $\Gamma_{GS}$, i.e. $\Gamma_{ES}\gg\Gamma_{GS}$. Then, we can read out the spin state as follows. At time $t$=0, we position the levels of both \ket{\:ES} and \ket{\:GS} far above $\mu_{res}$, so that one electron is energetically allowed to tunnel off the dot regardless of the spin state. Then, at a time $t=\tau$, where $\Gamma_{GS}^{-1}\gg \tau \gg \Gamma_{ES}^{-1}$, an electron will have tunneled off the dot with a very high probability if the state was \ket{\:ES}, but most likely no tunneling will have occurred if the state was \ket{\:GS}. Thus, the spin information is converted to charge information, and a measurement of the number of electrons on the dot reveals the original spin state.

A major advantage of this TR-RO scheme is that it does not rely on a large energy splitting between the spin states. Furthermore, it is robust against background charge fluctuations, since these cause only a small variation in the tunnel rates (of order $10^{-3}$ in Ref. \cite{FujisawaChargeNoise}). 
Finally, photon-assisted tunneling is not important since here tunneling is energetically allowed regardless of the initial spin state. Thus, we see that TR-RO can overcome the constraints of E-RO.

We first analyze the fidelity of the TR-RO theoretically using the error rates $\alpha$ and $\beta$ as defined in the diagram of Fig.~\ref{TR-ROFig1}c (inset). Here, $\alpha$ is the probability that one electron has tunneled even though the initial state was \ket{\:GS}, and $\beta$ the probability that no tunneling has occurred even though the initial state was \ket{\:ES}. The charge measurement itself is assumed to be perfect, and spin relaxation from \ket{\:ES} to \ket{\:GS} is modeled by a rate 1/$T_1$. 
We find analytically
\begin{eqnarray}
	\alpha\!&=&\!\!1-e^{-\Gamma_{GS} \cdot \tau}, \label{eqn:alpha}\\
	\beta\!&=&\!\!\frac{(1/T_1) e^{-\Gamma_{GS} \cdot \tau}+\left(\Gamma_{ES}\!-\!\Gamma_{GS}\right) e^{-(\Gamma_{ES} + 1/T_1) \cdot \tau}}{\Gamma_{ES}+1/T_1-\Gamma_{GS}}, \label{eqn:beta}
\end{eqnarray}
where $\tau$ is the time at which we measure the number of electrons $N$. The visibility of the read-out is $1\!-\!\alpha\!-\!\beta$.

In Fig.~\ref{TR-ROFig1}c we plot the visibility for the optimal value of $\tau$
as a function of $T_1$ and the ratio of the tunnel rates $\Gamma_{ES}/\Gamma_{GS}$. (Here, $\Gamma_{GS}$ is chosen to be 2.5~kHz, which is well within the bandwidth of our charge detection set up \cite{LievenAPL}.) We see that for $\Gamma_{ES}/\Gamma_{GS}=10$ and $T_1~\!=\!~0.5$~ms, the visibility is 65\%, equal to the visibility obtained with E-RO in Ref.~\cite{NatureReadout} for the same $T_1$. For $\Gamma_{ES}/\Gamma_{GS}>60$ and $T_1~\!=\!~0.5$~ms, the visibility of TR-RO exceeds 90\%.

The TR-RO can be used in a similar way if $\Gamma_{ES}$ is much \textit{lower} than $\Gamma_{GS}$. The visibility for this case can be calculated simply by replacing $\alpha$ and $\beta$ in Eqs.~\ref{eqn:alpha}-\ref{eqn:beta} with $1-\alpha$ and $1-\beta$ respectively. Due to the symmetry in the equations, this visibility is the same as for the case $\Gamma_{ES}\gg\Gamma_{GS}$ whenever the relaxation rate, which is the only asymmetric parameter, is not dominant.

The main ingredient necessary for TR-RO is a spin dependence in the tunnel rates. For a single electron, this spin dependence can be obtained in the Quantum Hall regime, where a high spin-selectivity is induced by the spatial separation of spin-resolved edge channels~\cite{Ciorga,CiorgaNDR}.
TR-RO can also be used for read-out of a two-electron dot, where the electrons are either in the spin-singlet ground state, denoted by \ket{\:S}, or in a spin-triplet state, denoted by \ket{\:T}. In \ket{\:S}, the two electrons both occupy the lowest orbital, but in \ket{\:T} one electron is in the first excited orbital. Since the wave function in this excited orbital has more weight near the edge of the dot \cite{LeoFewEl}, the coupling to the reservoir is stronger than for the lowest orbital. Therefore, the tunnel rate from a triplet state to the reservoir $\Gamma_T$ is much larger than the rate from the singlet state $\Gamma_S$, i.e. $\Gamma_T\gg\Gamma_S$ \cite{RonaldMoriond}. We use this spin-dependence in the following to experimentally demonstrate TR-RO for two electrons.

\begin{figure}[!t]
\includegraphics[width=3.4in]{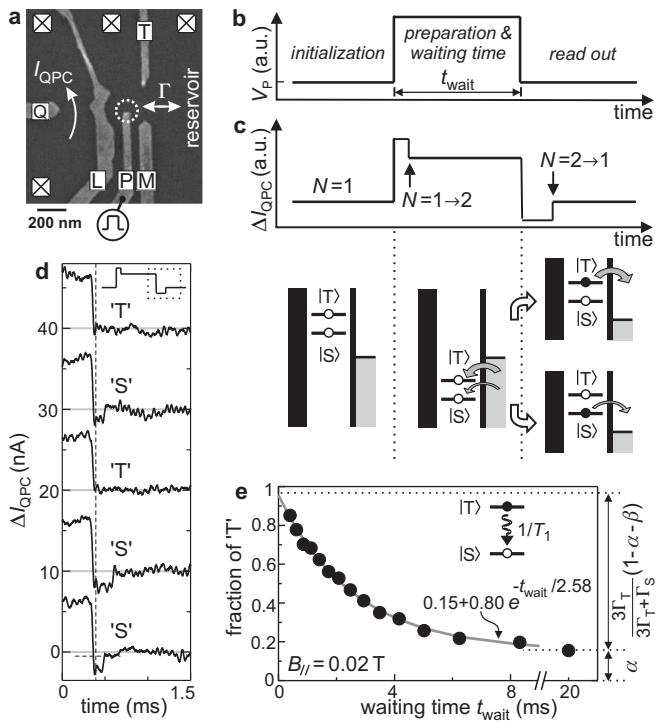}
\caption{Single-shot read-out of $N\!=\!2$ spin states. (a) Scanning electron micrograph of a device as used in the experiments. 
(b) Pulse waveform applied to gate \textit{P}. (c) Response of the QPC-current to the waveform of (b). Energy diagrams indicate the positions of the levels during the three stages. In the final stage, spin is converted to charge information due to the difference in tunnel rates for states \ket{\:S} and \ket{\:T}. (d) Real-time traces of $\Delta I_{QPC}$ during the last part of the waveform (dashed box in the inset), for $t_{wait}\!=0.8$~ms. At the vertical dashed line, $N$ is determined by comparison with a threshold (horizontal dashed line in bottom trace) and the spin state is declared $'T'$ or $'S'$ accordingly. (e) Fraction of $'T'$ as a function of waiting time at $B_{/\!/}\!$~=~0.02~T, showing a single-exponential decay with a time constant $T_1$ of 2.58~ms.
}
\label{TR-ROFig2}
\end{figure}

A quantum dot (white dotted circle in Fig.~\ref{TR-ROFig2}a) and a QPC are defined in a two-dimensional electron gas (2DEG) with an electron density of $4\cdot10^{15}$~m$^{-2}$), 60~nm below the surface of a GaAs/AlGaAs heterostructure from Sumitomo Electric, by applying negative voltages to gates $L$, $M$, $T$ and $Q$. Gate $P$ is used to apply fast voltage pulses. We completely pinch off the tunnel barrier between gates $L$ and $T$, so that the dot is only coupled to the reservoir on the right. The conductance of the QPC is tuned to about $e^2/h$, making it very sensitive to the number of electrons on the dot. A voltage bias of 0.8~mV induces a current through the QPC, $I_{QPC}$, of about 30~nA.

We tune the dot to the $N\!=\!1\!\leftrightarrow\!2$ transition in a small parallel field $B_{/\!/}$ of 0.02~T. Here, the energy difference between \ket{\:T} and the ground state \ket{\:S}, $E_{ST}$, is about 1~meV. From measurements of the tunnel rates \cite{JeroAPL}, we estimate the ratio $\Gamma_T/\Gamma_S$ to be on the order of 20. A similar ratio was found previously in transport measurements on a different device \cite{RonaldMoriond}. As can be seen in Fig.~\ref{TR-ROFig1}c, for $T_1~\!>\!1\!~$ms this permits a read-out visibility $\!>\!80$\%.

We implement the TR-RO by applying voltage pulses as depicted in Fig.~\ref{TR-ROFig2}b to gate $P$. Figure \ref{TR-ROFig2}c shows the expected response of $I_{QPC}$ to the pulse, together with the level diagrams in the three different stages. Before the pulse starts, there is one electron on the dot. Then, the pulse pulls the levels down so that a second electron can tunnel onto the dot ($N\!=\!1\!\rightarrow\!2$), forming either a singlet or a triplet state with the first electron.
The probability that a triplet state is formed is given by $3\Gamma_T/(\Gamma_S + 3\Gamma_T)$, where the factor of 3 is due to the degeneracy of the triplets. After a variable waiting time $t_{wait}$, the pulse ends and the read-out process is initiated, during which one electron can leave the dot again. The rate for tunneling off depends on the two-electron state, resulting in the desired spin-to-charge conversion. The QPC is used to detect the number of electrons on the dot. Due to the direct capacitive coupling of gate $P$ to the QPC channel, $\Delta I_{QPC}$ follows the pulse shape. Tunneling of an electron on or off the dot gives an additional step in $\Delta I_{QPC}$ \cite{LievenAPL,NatureReadout,EnsslinAPL}, as indicated by the arrows in Fig. \ref{TR-ROFig2}c.

Now, $\Gamma_S$ is tuned to 2.5~kHz, and $\Gamma_T$ is therefore $\approx\:$50~kHz. In order to achieve a good signal-to-noise ratio in $I_{QPC}$, the signal is sent through an external 20~kHz low-pass filter. As a result, many of the tunnel events from \ket{\:T} will not be resolved, but the tunneling from \ket{\:S} should be clearly visible.

Figure \ref{TR-ROFig2}d shows several traces of $\Delta I_{QPC}$, from the last part (300 $\mu$s) of the pulse to the end of the read-out stage (see inset), for a waiting time of 0.8~ms. In some traces, there are clear steps in $\Delta I_{QPC}$, due to an electron tunneling off the dot. In other traces, the tunneling occurs faster than the filter bandwidth. In order to discriminate between \ket{\:S} and \ket{\:T}, we first choose a read-out time $\tau$ (indicated by a vertical dashed line in Fig. \ref{TR-ROFig2}d) and measure the number of electrons on the dot at that time by comparing $\Delta I_{QPC}$ to a threshold value (as indicated by the horizontal dashed line in the bottom trace of Fig. \ref{TR-ROFig2}d). If $\Delta I_{QPC}$ is below the threshold, it means $N\!=\!2$ and we declare the state $'S'$. If $\Delta I_{QPC}$ is above the threshold, it follows that $N\!=\!1$ and the state is declared $'T'$. Our method for determining the optimal threshold value and $\tau$ is explained below.

To  verify that $'T'$ and $'S'$ indeed correspond to the spin states \ket{\:T} and \ket{\:S}, we change the relative occupation probabilities by varying the waiting time. The probability that the electrons are in \ket{\:T}, $P_T$, decays exponentially with the waiting time: $P_{T}(t)=P_{T}(0)\,e^{-t_{wait}/T_1}$. Therefore, as we make the waiting time longer, we should observe an exponential decay of the fraction of traces that are declared $'T'$.

We take 625 traces similar to those in Fig.~\ref{TR-ROFig2}d for each of 15 different waiting times. Note that the two-electron state is formed on a timescale (of order 1/$\Gamma_T$) much shorter than the shortest $t_{wait}$ used (400~$\mu$s). To find the optimal read-out parameters, we scan a wide range of read-out times and threshold values using a computer program. For each combination of these two parameters, the program determines the fraction of traces declared $'T'$ for each of the waiting times, and fits the resulting data with a single exponential decay $A\,e^{-t_{wait}/T_1}+\alpha$. The prefactor $A$ is given by $3\Gamma_T/(\Gamma_S+3\Gamma_T)\! \times\!  (1\!-\!\alpha\!-\!\beta)$. We see that $A$ is proportional to the read-out visibility, and therefore the optimal read-out parameters can be determined simply by searching for the highest value of $A$. Here, we find the optimal values to be -0.4~nA for the threshold and 70~$\mu$s for $\tau$ (corresponding to $t$ = 370~$\mu$s in Fig.~\ref{TR-ROFig2}d), and use these in the following.

In Fig.~\ref{TR-ROFig2}e, we plot the fraction of traces declared $'T'$ as a function of $t_{wait}$. We see that the fraction of $'T'$ decays exponentially, showing that we can indeed read out the two-electron spin states. A fit to the data yields a triplet-to-singlet relaxation time $T_1\!=\!(2.58\pm0.09)$~ms, which is more than an order of magnitude longer than the lower bound found in Ref.~\cite{FujisawaNature}. As indicated on the right side of Fig.~\ref{TR-ROFig2}e, we can also extract $\alpha$ and $\beta$ from the data. We find $\alpha\!~=\!~0.15$ and $\beta\!=\!~0.04$ (taking $\Gamma_T/\Gamma_S~=~20$). The single-shot visibility is thus 81\%. These numbers agree well with the values predicted by the model ($\alpha=0.14$, $\beta=0.05$, visibility$=81$\%), as indicated by the diamond in Fig.~\ref{TR-ROFig1}c. Note that, since the visibility is insensitive to $\tau$ near the optimal value, it is not significantly reduced by the finite bandwidth of the charge measurement.

As an extra check of the read-out, we have also applied a modified pulse where during the preparation only the singlet state is energetically accessible. Here, the read-out should ideally always yield $'S'$, and therefore the measured probability for finding $'T'$ directly gives us $\alpha$. We find a fraction of $'T'$ of 0.16, consistent with the value of $\alpha$ obtained from the fit. This again confirms the validity of the read-out method.

\begin{figure}[!t]
\includegraphics[width=3.4in]{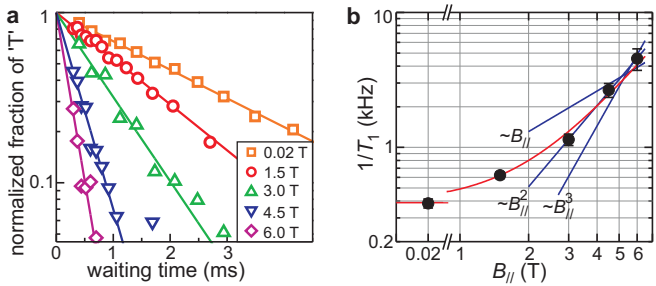}
\caption{Triplet-to-singlet relaxation as a function of $B_{/\!/}$. (a) Normalized fraction of $'T'$ vs. $t_{wait}$ for different values of $B_{/\!/}$. (b) Triplet-to-singlet relaxation rate 1/$T_1$ as a function of $B_{/\!/}$. The data is fit to with a second-order polynomial (see text). For comparison, lines with linear, quadratic and cubic $B_{/\!/}$-dependences are shown.
}
\label{TR-ROFig3}
\end{figure}
We further study the relaxation between triplet and singlet states by repeating the measurement of Fig.~\ref{TR-ROFig2}e at different magnetic fields $B_{/\!/}$. Figure~\ref{TR-ROFig3}a shows the decay of the fraction of $'T'$, normalized to the fraction of $'T'$ at $t_{wait}\!=\!0$,
on a logarithmic scale. The data follow a single-exponential decay at all fields. Figure~\ref{TR-ROFig3}b shows the relaxation rate 1/$T_1$ as a function of $B_{/\!/}$. The dominant relaxation mechanisms for large values of $E_{ST}$ are believed to originate from the spin-orbit interaction \cite{VitalyST,KhaetskiiST}, but to our knowledge the case of an in-plane magnetic field has not been treated yet. A second-order polynomial fit to the data yields $1/T_1$~[kHz]= $(0.39\pm0.03)+(0.10\pm0.02)\cdot B_{/\!/}^2$~[T], with a negligible linear term.

\begin{figure}[!t]
\includegraphics[width=3.4in]{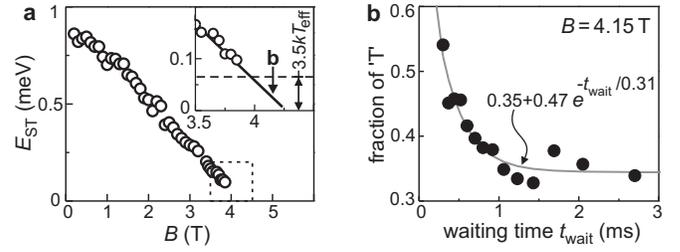}
\caption{Single-shot read-out of nearly degenerate states. (a) Singlet-triplet energy difference $E_{ST}$ as a function of magnetic field $B$, applied under a 45 degree angle with the 2DEG. Inset: zoom-in of the region inside the dashed square. For $B~>$~3.9~T, $E_{ST}$ is smaller than the effective electron temperature. (b) Single-shot read-out at $B$~=~4.15~T. This field value is indicated with `b' in the inset of (a).
}
\label{TR-ROFig4}
\end{figure}

Finally, we show that the TR-RO can still be used when \ket{\:S} and \ket{\:T} are almost degenerate. By mounting the device under a 45 degree angle with respect to the magnetic field axis, we can tune $E_{ST}$ through zero \cite{LeoFewEl}. In Fig.~\ref{TR-ROFig4}a we plot $E_{ST}$ as a function of $B$, extracted from pulse spectroscopy measurements \cite{JeroAPL}. In these measurements, transitions are broadened both by the electron temperature in the reservoir and by fluctuations in the dot potential. We model these two effects by one effective electron temperature $T_{e\!f\!f}$. For $E_{ST}$ smaller than about 3.5~$kT_{e\!f\!f}$, the energy splitting can not be resolved. As in previous transport and pulse spectroscopy measurements, we find here 3.5~$kT_{e\!f\!f}\approx~\!60~\mu$eV (see inset of Fig.~\ref{TR-ROFig4}a), and therefore it is impossible to use the E-RO method beyond $B\!~\!\approx\!~\!$3.9~T. From extrapolation of the data, we find that the singlet-triplet ground state transition occurs at (4.25$\!~\!\pm\!~\!0.05)\!~$T.

We tune $B$ to 4.15 T (see inset of Fig.~\ref{TR-ROFig4}a), so that we are very close to the degeneracy point, but still certain that \ket{\:S} is the ground state. Figure~\ref{TR-ROFig4}b shows the result of the read-out measurement
at this field.
Again, an exponential decay of the fraction of $'T'$ is observed, with a $T_1$ of ($0.31\!~\pm\!~0.07$)~ms. This demonstrates that even when the energy splitting $E_{ST}$ is too small to resolve, we can still read out the spin states using TR-RO. In future measurements, we plan to apply the tunnel-rate-selective read-out to detect relaxation and coherent manipulation of a single electron spin.

We thank V. Golovach, S.I. Erlingsson and D. Loss for useful discussions. This work was supported by FOM, NWO, the DARPA-QUIST program, the ONR and the EU-RTN network on spintronics.

\end{document}